# One-Log Call Iterative Solution of the Colebrook Equation for Flow Friction Based on Padé Polynomials

**Pavel Praks** [1] and Dejan Brkić [2]

[1] European Commission, DG Joint Research Centre (JRC), Directorate C: Energy, Transport and Climate, Unit C3: Energy Security, Distribution and Markets, Via Enrico Fermi 2749, 21027 Ispra (VA), Italy and IT4Innovations National Supercomputing Center, VŠB - Technical University Ostrava, 17. listopadu 2172/15, 708 00 Ostrava, Czech Republic; ORCID id: 0000-0002-3913-7800, Pavel.Praks@ec.europa.eu, Pavel.Praks@vsb.cz (P.P.)

[2] European Commission, DG Joint Research Centre (JRC), Directorate C: Energy, Transport and Climate, Unit C3: Energy Security, Distribution and Markets, Via Enrico Fermi 2749, 21027 Ispra (VA), Italy; ORCID id: 0000-0002-2502-0601, dejanbrkic0611@gmail.com, dejanrgf@tesla.rcub.bg.ac.rs (D.B.)

**Abstract:** The eighty years old empirical Colebrook function $\xi$ widely used as an informal standard for hydraulic resistance relates implicitly the unknown flow friction factor $\lambda$, with the known Reynolds number $Re$ and the known relative roughness of a pipe inner surface $\varepsilon^*$; $\lambda = \xi(Re, \varepsilon^*, \lambda)$. It is based on logarithmic law in the form that captures the unknown flow friction factor $\lambda$ in a way from which it cannot be extracted analytically. As an alternative to the explicit approximations or to the iterative procedures that require at least a few evaluations of computationally expensive logarithmic function or non-integer powers, this paper offers an accurate and computationally cheap iterative algorithm based on Padé polynomials with only one *log*-call in total for the whole procedure (expensive *log*-calls are substituted with Padé polynomials in each iteration with the exception of the first). The proposed modification is computationally less demanding compared with the standard approaches of engineering practice, but does not influence the accuracy or the number of iterations required to reach the final balanced solution.

**Keywords:** Colebrook equation; Colebrook-White; flow friction; iterative procedure; logarithms; Padé polynomials; hydraulic resistances; turbulent flow; pipes; computational burden

## 1. Introduction

The empirical Colebrook equation (Colebrook 1939) implicitly relates the unknown flow friction factor $\lambda$ with the known Reynolds number $Re$ and the know relative roughness of inner pipe surface, $\varepsilon^*$; $\lambda = \xi(Re, \varepsilon^*, \lambda)$, where $\xi$ is functional symbol, Equation (1).

$$\frac{1}{\sqrt{\lambda}} = -2 \cdot log_{10}\left(\frac{2.51}{Re} \cdot \frac{1}{\sqrt{\lambda}} + \frac{\varepsilon^*}{3.71}\right) \tag{1}$$

In Equation (1) $Re$ is Reynolds number; $4000 < Re < 10^8$, $\varepsilon^*$ is relative roughness of inner pipe surface; $0 < \varepsilon^* < 0.05$, and $\lambda$ is Darcy flow friction factor; $0.0064 < \lambda < 0.077$ (all three quantities are dimensionless).

The Colebrook equation is based on experiments performed by Colebrook and White in 1937 with the flow of air through a set of artificially roughened pipes (Colebrook and White 1937). The accuracy of this eighty year old equation is disputed many times but it is still accepted in engineering practice as an informal standard for hydraulic resistance. Therefore to repeat results and for comparisons, it is required to solve the Colebrook equation accurately. Numerous evaluations of flow friction factor such as in the case of complex networks of pipes pose extensive burden for computers, so not only an accurate but also a simplified solution is required. Calculation of complex water or gas distribution networks (Brkić 2009, Brkić 2011ab, Praks et al. 2015, Praks et al. 2017) which requires few evaluations of logarithmic function for each pipe, presents a significant and extensive burden which available computer resources hardly can easily manage (Clamond 2009, Giustolisi et al. 2011, Danish et al. 2011, Winning and Coole 2013, Vatankhah 2018).

The Colebrook equation is based on logarithmic law where the unknown flow friction factor $\lambda$ is given implicitly, i.e., it appears on both sides of Equation (1) in form $\lambda = \xi(Re, \varepsilon^*, \lambda)$, from which it cannot be extracted analytically; an exception is through the Lambert $W$-function (Keady 1998, Sonnad and Goudar 2004, Brkić 2011cd, Brkić 2012ab, Biberg 2017, Brkić 2017a). The common way to solve it is to guess an initial value $\lambda_0$ for friction factor and then to try to balance it using the iterative algorithm (Brkić 2017b) which needs to be terminated after the certain number of iterations when the final balanced value $\lambda_n$ is reached. As an alternative to the iterative procedure, numerous approximate formulas are available (Gregory and Fogarasi 1985, Zigrang and Sylvester 1985, Brkić 2011e, Brkić 2012c, Brkić and Ćojbašić 2017, Pimenta et al. 2018). Usually, more complex approximations are more accurate, but also more computationally expensive because they contain at least a few logarithmic expressions and/or terms with non-integer powers which require use of demanding algorithms (namely an evaluation of fractional exponential and natural logarithm) to be evaluated in processor units of computers and to be stored in registers (Clamond 2009, Giustolisi et al. 2011, Danish et al. 2011, Winning and Coole 2013, Vatankhah 2018, Sonnad and Goudar 2004).

The presented scheme for solving Colebrook equation requires only one single call of the logarithmic function in respect to the whole iterative procedure. It is equally accurate as a standard iterative approach and does not require additional iterations to reach the same accuracy. Instead of computationally expensive logarithmic function, its Padé polynomial equivalent (Baker and Graves-Morris 1996) is used in all iterations, exception the first. The Padé approximant is the approximation of a function by a rational algebraic fraction where both the numerator and the denominator are polynomials. Because these rational functions only use the elementary arithmetic operations, they can be evaluated numerically very easily. In the computer environment, they required less basic floating-point operations compared with the logarithmic function (Kropa 1978, Rising 2007, Pineiro et al. 2004, Al-Mohy 2012).

The presented simplified iterative method can be profitable for future computing software in terms of having a high level of accuracy and speed with a decreased computational burden.

## 2. Evaluation of Logarithmic Function through Padé Polynomials

Basic floating-point operations such as addition and multiplication are carried out directly in the Central Processor Unit (CPU) while logarithmic functions, exponents or square roots require expensive operations based on more complex algorithms (Kropa 1978, Rising 2007, Pineiro et al. 2004, Al-Mohy 2012). In addition to logarithms and non-integer powers, Biberg (2017) adds also division in the group of more costly functions for evaluation; addition, subtraction and multiplication are low-cost operations according to Biberg. Evaluated with various compilers and executed on various platforms, integer addition, subtraction, or multiplication requires less than 1 floating-point operation, float addition, subtraction, or multiplication about 1, float division 2–6, integer division 4–10, square root 5–20, while functions *sin*, *cos*, *tan*, as well *log* and *exp* functions 10–40 floating-point operations. Winning and Coole (2015) report average time for 100 million operation in seconds and relative effort, respectively as follows: addition 23.40s and 1, subtraction 27.50s and 1.18, division 31.70s and 1.35, multiplication 36.20s and 1.55, squared 51.10s and 2.18, square root 53.70s and 2.29, cubed 55.58s and 2.38, natural log 63.00s and 2.69, cubed root 63.40s and 2.71, fractional exponential 77.60s and 3.32, and log to 10-base 78.80s and 3.37.

Regarding the Colebrook equation, in order to simplify the iterative procedure which is in common use in engineering practice, the logarithmic function is replaced with its relevant Padé polynomial equivalent in all iterations, with exception to the first. Padé polynomials can accurately approximate logarithmic function only in a limited domain. For example, if it is known that $log_{10}(100) = 2$, calculation of for example $log_{10}(90)$ can be evaluated using the fact that $log_{10}(100/90) = log_{10}(100) - log_{10}(90) \rightarrow log_{10}(90) = log_{10}(100) - log_{10}(100/90)$ where $100/90 \approx 1.111$ is close to 1. Logarithmic function can be replaced by its Padé polynomial equivalent very accurately in a limited domain, and therefore instead of $log_{10}(1.111)$, already calculated $log_{10}(100) = 2$ and Padé polynomial which is accurate around 1 for argument $z = 1.1111$ can be used to calculate $log_{10}(90)$.

Because of linearization of the unknown parameter $\lambda$, a more suitable form of the Colebrook equation for computation is $x = -2 \cdot log_{10}\left(\frac{2.51 \cdot x}{Re} + \frac{\varepsilon^*}{3.71}\right)$, where $x = \frac{1}{\sqrt{\lambda}}$. The argument of logarithmic function in the Colebrook equation is $y = \frac{2.51 \cdot x}{Re} + \frac{\varepsilon^*}{3.71}$ where only evaluation through its native logarithmic form $log_{10}(y)$ need go only in the first iteration where further evaluation can go through the appropriate Padé polynomial which is accurate for its argument $z$ around 1, knowing that $z_{01} = \frac{y_0}{y_1}$, $z_{02} = \frac{y_0}{y_2}$, $z_{03} = \frac{y_0}{y_3}$, etc. or $z_{01} = \frac{y_0}{y_1}$, $z_{12} = \frac{y_1}{y_2}$, $z_{23} = \frac{y_2}{y_3}$, etc. in the case of the Colebrook equation it is always near 1; $z \approx 1$. Evaluation of 10-base logarithmic function in many computing languages goes through natural logarithm where $log_{10}(z) = \frac{ln(z)}{ln(10)}$ and where $ln(10) \approx 2.302585093$ is constant, and therefore the Padé polynomials that approximate accurately $ln(z)$ for $z \approx 1$ are shown; Equations (2-7). Padé polynomials of different orders can be used for approximation of $ln(z)$, here all accurate for arguments $z$ close to 1; $z \approx 1$. As the expansion point $z_0 = 1$ is a root of $ln(z)$, the accuracy of the Padé approximant decreases. Setting the OrderMode option in Matlab padé command to relative compensates for the decreased accuracy. Thus, here, the Pade approximant of $(m, n)$ order uses the form $ln(z) \approx \frac{(z-z_0) \cdot (a_0 + \alpha_1 \cdot (z-z_0) + \cdots + \alpha_1 \cdot (z-z_0)^m)}{1 + \beta_1 \cdot (z-z_0) + \cdots + \beta_n \cdot (z-z_0)^n}$, where $\alpha$ and $\beta$ are coefficients (the coefficients of the polynomials need not be rational numbers). For example, Padé polynomial of order (2, 3) is with polynomial of order 2 in numerator and of order 3 in denominator; Equation (6). Of course, low order formulas are simpler, but they have larger errors than high order formulas and vice versa. As can be seen from Figure 1, even very simple form of Padé polynomials (1,2) and (2,1) are of high accuracy in respect of domain of interest for solving the Colebrook equation which is $z \approx 1$; $z \in [0.9, 1.1]$. Horner algorithm transforms polynomials into a computationally efficient form and therefore, Horner nested polynomial representations of the Padé polynomials of different orders for $ln(z)$ where $z \to 1$ are shown here; Equations (2–7). Higher order of Padé approximants are more accurate, but more complex.

Order (1,1):

$$ln(z) \approx \frac{z \cdot (z+4) - 5}{4 \cdot z + 2} \tag{2}$$

Order (1,2):

$$ln(z) \approx \frac{3 \cdot (z-1) \cdot (z+1)}{z \cdot (z+4) + 1} \tag{3}$$

Order (2,1):

$$ln(z) \approx \frac{-z \cdot (z \cdot (z-9) - 9) - 17}{18 \cdot z + 6} \tag{4}$$

Order (2,2):

$$ln(z) \approx \frac{z \cdot (z \cdot (z+18) - 9) - 10}{z \cdot (9 \cdot z + 18) + 3} \tag{5}$$

Order (2,3):

$$ln(z) \approx \frac{(z-1) \cdot (11 \cdot z^2 + 38 \cdot z + 11)}{3 \cdot (z^3 + 9 \cdot z^2 + 9 \cdot z + 1)} \tag{6}$$

Order (3,2):

$$ln(z) \approx \frac{z \cdot (z \cdot (11 \cdot z + 27) - 27) - 11}{z \cdot (z \cdot (3 \cdot z + 27) + 27) + 3} \tag{7}$$

In Equations (2-7), $z$ is from $z_{01} = \frac{y_0}{y_1}$, $z_{02} = \frac{y_0}{y_2}$, $z_{03} = \frac{y_0}{y_3}$, etc., or $z_{01} = \frac{y_0}{y_1}$, $z_{12} = \frac{y_1}{y_2}$, $z_{23} = \frac{y_2}{y_3}$, etc.; and $y = \frac{2.51 \cdot x}{Re} + \frac{\varepsilon^*}{3.71}$.

Relative error introduced by them; Equations (2-5) compared with $ln(z)$ is shown in Figure 1 and for Equation (6) in Table 1. The relative error of Padé approximants (2,2) for $z \approx 1$ of $ln(z)$ is negligible for $0.8 < z < 1.2$. Thus, relative error of the used Padé approximants (2,3) of $ln(z)$ in the proposed iterative procedure is even more negligible and therefore it is not presented in Figure 1, but is available in Table 1.

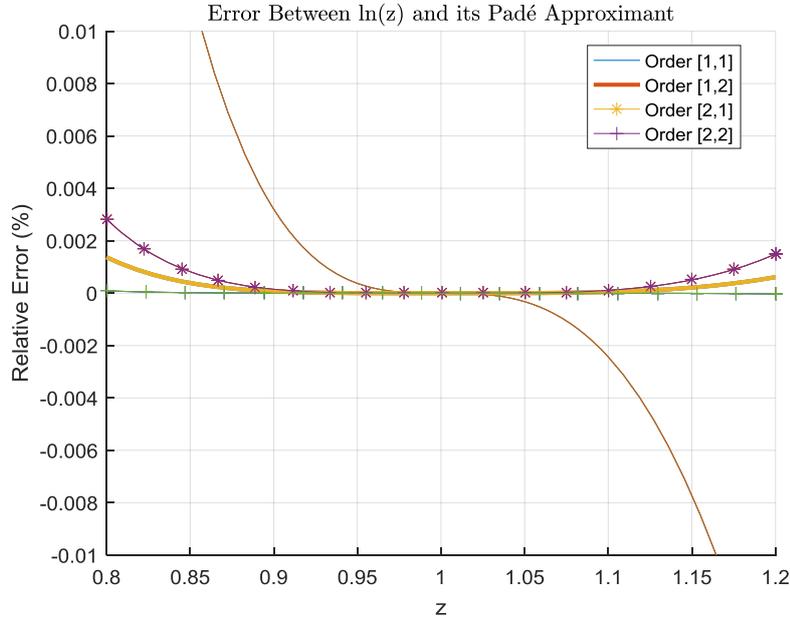

Figure 1. Relative error between ln(z) and its Padé approximants accurate for $z \approx 1$

Table 1: Relative error in % of Padé approximant (2,3) for z in interval [0.6; 1,6]

| z | $log_{10}(z) = \frac{ln(z)}{ln(10)}$ | Padé approximants (2,3) | Relative error % |
|---|---|---|---|
| 0.6 | -0.22184875 | -0.221847398 | $6.1 \cdot 10^{-4}$% |
| 0.65 | -0.187086643 | -0.187086228 | $2.2 \cdot 10^{-4}$% |
| 0.7 | -0.15490196 | -0.154901848 | $7.2 \cdot 10^{-5}$% |
| 0.75 | -0.124938737 | -0.124938712 | $2.0 \cdot 10^{-5}$% |
| 0.8 | -0.096910013 | -0.096910009 | $4.4 \cdot 10^{-6}$% |
| 0.85 | -0.070581074 | -0.070581074 | $6.6 \cdot 10^{-7}$% |
| 0.9 | -0.045757491 | -0.045757491 | $4.9 \cdot 10^{-8}$% |
| 0.95 | -0.022276395 | -0.022276395 | $6.5 \cdot 10^{-10}$% |
| 1 | 0 | 0 | 0% |
| 1.05 | 0.021189299 | 0.021189299 | $4.8 \cdot 10^{-10}$% |
| 1.1 | 0.041392685 | 0.041392685 | $2.7 \cdot 10^{-8}$% |
| 1.15 | 0.06069784 | 0.06069784 | $2.7 \cdot 10^{-7}$% |
| 1.2 | 0.079181246 | 0.079181245 | $1.3 \cdot 10^{-6}$% |
| 1.25 | 0.096910013 | 0.096910009 | $4.4 \cdot 10^{-6}$% |
| 1.3 | 0.113943352 | 0.113943339 | $1.2 \cdot 10^{-5}$% |
| 1.35 | 0.130333768 | 0.130333735 | $2.6 \cdot 10^{-5}$% |
| 1.4 | 0.146128036 | 0.146127961 | $5.1 \cdot 10^{-5}$% |
| 1.45 | 0.161368002 | 0.161367854 | $9.2 \cdot 10^{-5}$% |
| 1.5 | 0.176091259 | 0.176090987 | $1.5 \cdot 10^{-4}$% |
| 1.55 | 0.190331698 | 0.190331231 | $2.5 \cdot 10^{-4}$% |
| 1.6 | 0.204119983 | 0.204119223 | $3.7 \cdot 10^{-4}$% |

To illustrate the complexity of computing in modern computers it should be noted that even such a relatively simple equation such as Colebrook's can make a numerical problem in computer registers due to overflow error. Its transformed version in term of the Lambert $W$-function can give such large numbers for some pairs of the Reynolds number Re and the relative roughness of inner pipe surface $e^*$ which are from the practical domain of applicability in engineering practice which cannot be stored in 32- or 64-bit registers of modern computers (Sonnad and Goudar 2004, Brkić 2012a).

## 3. Initial Starting Point for the Proposed Iterative Method

In case of the Colebrook equation, practical experience shows that trying to get a good initial starting point $x_0$ has limited value until it is chosen in the domain of applicability of the equation which is $3.68 < x < 12.47$. Every initial starting point $x_0$ chosen from the domain of applicability of the Colebrook equation will lead to the final accurate solution surely, with the only difference that in some cases more additional iterations would be needed. Usually, with the initial guess $x_0$ that is close to the exact solution, the iterative procedure converges to it in five or fewer iterations. To date, cases which lead to divergence, fluctuation, or convergence to a possible far away solution outside of the practical domain of applicability of the Colebrook equation are not known. In of the proposed approach, a good starting point should be chosen within the domain of applicability of the Colebrook equation and should not contain any logarithmic function and/or non-integer power term.

A number of options to choose an optimal starting point $x_0$ are considered: 1) special case of the Colebrook equation when $Re \to \infty$, 2) integration of the Colebrook equation, 3) explicit approximations of the Colebrook equation, and 4) fixed value.

1. The common approach is to choose an initial starting point from the zone of fully developed turbulent rough hydraulic flow $x_0 = -2 \cdot \log_{10}\left(\frac{e^*}{3.71}\right)$, because in this special case of the Colebrook equation where $Re \to \infty$, the equation is in explicit form with respect to $x$; $x_0 = \xi(e^*)$, where $\xi$ is functional symbol (Brkić 2017). Here the goal is to avoid use of logarithmic functions and therefore, this starting point is not suitable.

2. An efficient procedure for finding a sufficiently good initial starting point $x_0$ is proposed by Yun (2008) in the integral form; Equation (8):

$$x_0 = \frac{1}{2} \cdot \left\{ a + b + sgn(F(a)) \cdot \int_a^b tanh(F(x)) \, dx \right\} \tag{8}$$

In Equation (8), $F = x - \xi(x) = 0$, $\xi$ represents the Colebrook equation, $a$ is the lower while $b$ is the upper limit from which an initial starting point $x_0$ should be chosen; $a = 3.68$ and $b = 12.47$ because the domain of applicability of the Colebrook equation that is between 3.68 and 12.47 in respect to $x$, $sgn$ is signum function: if $F(a) > 0 \to sgn(F(a)) = 1$, $F(a) = 0 \to sgn(F(a)) = 0$, and $F(a) < 0 \to sgn(F(a)) = -1$, while $tanh$ is hyperbolic tangent which is defined through the exponential function $e^x$ with non-integer power $x$ the use of which is as computationally expensive as the use of the logarithmic function and which therefore cannot be recommended.

3. Every explicit approximation of the Colebrook equation (Gregory and Fogarasi 1985, Zigrang and Sylvester 1985, Brkić 2011, Brkić and Ćojbašić 2017, Pimenta et al. 2018); $x \approx \varsigma(Re, e^*)$, where $\varsigma$ is the functional symbol, can be used to choose an initial starting point $x_0$. On the other hand, almost all available approximations contain logarithmic or/and terms with non-integer powers, which makes them unsuitable for use in the developed approach. On the other hand, having previous experience with training Artificial Neural Networks (ANN) to simulate the Colebrook equation (Özger and Yıldırım 2009, Brkić and Ćojbašić 2016, Bardestani et al. 2017), i.e. to use ability of artificial intelligence to simulate the Colebrook equation not knowing its logarithmic nature but only knowing raw input and corresponding output datasets $\{Re, e^*\} \to \{x\}$, a computationally cheap explicit approximation of the Colebrook equation is developed through genetic programming (Giustolisi and Savić 2006, Ćojbašić and Brkić 2013, Brkić 2014, Brkić and Ćojbašić 2016). The developed approximation is computationally efficient because of its polynomial structure; Equation (9):

$$x_0 = 5.05 - 30.73 \cdot \varepsilon^* + \frac{3.4 \cdot Re + \frac{Re^2}{469647.7}}{46137.9 + Re + \frac{Re^2}{3250657.6} + \frac{\varepsilon^* \cdot Re^2}{515.25}} \tag{9}$$

Eureqa [computer software] by Nutonian, Inc., Boston, MA. (Schmidt and Lipson 2009, Dubčáková 2011) is used to generate Equation (9). The Eureqa-polynomial approximation; Equation (9) has up to 40% relative error, but it is very cheap and sufficiently accurate to serve as an initial starting point $x_0$.

4. Extensive tests over the domain of applicability of the Colebrook equation shows that one fixed value also can be used as the initial starting point $x_0$ for the iterative procedure in all cases. Results indicate that the proposed Padé approach works in all cases, as the argument $z$ of $ln(z)$ is always close to one. When Equation (9) is used, values of $z$ are within the range 0.91-1.05. Moreover, for the most pairs of the Reynolds number $Re$ and the relative roughness of inner pipe surfaces $\varepsilon^*$ which are in the domain of applicability, the initial starting point $x_0 = 7.273124147$ requires the least number of iterations.

To avoid using a computationally expensive logarithmic function in the initial stage of the iterative procedure, the recommendation is to start calculation with fixed-value starting point $x_0 = 7.273124147$ or to use a polynomial expression; Equation (9). Power-law formulas from Russian practice which do not contain non-integer powers also can be used (Альтшуль 1982).

## 4. Proposed Iterative Method

The Colebrook equation is usually solved iteratively using the Newton-Raphson method (Ypma 1995, Abbasbandy 2003) or even more using a simplified Newton-Raphson method known as the fixed-point method (Brkić 2017b). Recently, hybrid three-point methods have been proposed (Brkić and Praks 201x).

Here is presented an adjusted very accurate, fast and computationally cheap version of the Newton-Raphson method suitable for the Colebrook equation in which the logarithmic function is replaced after the first iteration with the Padé approximant in polynomial form (Baker and Graves-Morris 1996).

Knowing that the Colebrook equation is based on logarithmic law (Colebrook and White 1937, Colebrook 1939), the achievement with this simplified approach is more significant. Numerical examples are shown in Section 5 of this paper.

*Iteration 0:*

In order to avoid use of computationally expensive logarithmic functions or functions with non-integer powers, a required initial starting point $x_0$ should be chosen using recommendations from Section 3 of this paper; points 3 or 4.

*Iteration 1:*

Having provided an initial starting point $x_0$, new value $x_1$ can be calculated using Equation (10):

$$x_1 = x_0 - \frac{F(x_0)}{F'(x_0)} \tag{10}$$

In Equation (10), $F(x)$ represents the Colebrook equation $x = \xi(x)$ which needs to be in suitable form, $F = x - \xi(x) = 0$; Equation (11):

$$F(x_0) = x_0 + 2 \cdot \log_{10}(y_0) = 0 \tag{11}$$

In Equation (11), $y_0 = \frac{2.51 \cdot x_0}{Re} + \frac{\varepsilon^*}{3.71}$ which will be used also in the next iteration (in an additional variant of the proposed method $y_0$ is used in all subsequent iterations), while in Equation (10), the first derivative of $F$ in respect to $x$; $F'(x)$ is from Equation (12):

$$F'(x_0) = 1 + \frac{2 \cdot 2.51}{2.302585093 \cdot Re \cdot \left( \frac{100 \cdot \varepsilon^*}{371} + \frac{2.51 \cdot x_0}{Re} \right)} \tag{12}$$

In Equation (12), $ln(10) \approx 2.302585093$ is with constant value, and therefore only $log_{10}(y_0)$ from Equation (11) requires evaluation of the logarithmic function.

In many programming languages evaluation of logarithmic function of any base is processed by natural logarithm (Vatankhah 2018). Change of 10-base logarithm from the Colebrook equation to e-based natural logarithm where $e \approx 2.718$ and where $ln(10) \approx 2.302585093$ is implemented as $log_{10}(z) = \frac{ln(z)}{ln(10)}$.

***Iteration 2:***

New value $x_2$ should be calculated using Equation (13):

$$x_2 = x_1 - \frac{F(x_1)}{F'(x_1)} \tag{13}$$

In Equation (13), $F(x_1)$ is not calculated by $log_{10}(y_1)$, where $y_1 = \frac{2.51 \cdot x_1}{Re} + \frac{\varepsilon^*}{3.71}$, but using Padé polynomial replacement for logarithmic function which is accurate for $z \to 1$ and using the already calculated value of $log_{10}(y_0)$ from the previous iteration; Equation (14):

$$F(x_1) = x_1 + 2 \cdot log_{10}(y_0) - \frac{(z_{01} - 1) \cdot (11 \cdot z_{01}^2 + 38 \cdot z_{01} + 11)}{2.302585093 \cdot (3 \cdot z_{01}^3 + 9 \cdot z_{01}^2 + 9 \cdot z_{01} + 1)} \tag{14}$$

In Equation (14), $log_{10}(y_0) - \frac{(z_{01}-1) \cdot (11 \cdot z_{01}^2 + 38 \cdot z_{01} + 11)}{2.302585093 \cdot (3 \cdot z_{01}^3 + 9 \cdot z_{01}^2 + 9 \cdot z_{01} + 1)} \equiv log_{10}(y_1)$, $2.302585093 \approx ln(10)$, and $z_{01} = \frac{y_0}{y_1}$. In the first iteration, $log_{10}(y_0)$ is already known; Equation (11). The Padé polynomial used in Equation (14) is of order (2,3) which means that the polynomial in the numerator is of the order of 2 while in the denominator of order 3. The Padé polynomials are also known as Padé approximants and here the maximal relative error of the polynomial expression term in Equation (14) in domain $z \varepsilon [0.6, 1.6]$; $z \to 1$ is minor as shown in Table 1. Value of $z$ for the procedure shown in practice is $z \varepsilon [0.9, 1.1]$ and therefore the error of the used Padé approximant can be neglected in the case shown.

The first derivative $F'(x_1)$ does not contain any logarithmic functions and should be evaluated using Equation (12), where $x_0$ should be replaced with the new value $x_1$ or knowing that the value of the derivative does not change significantly between two iterations, $F'(x_0)$ can be reused in all subsequent iterative cycles. Even knowing that the value of the first derivate in the procedure shown is always near 1; for rough calculations it can be assumed that $F'(x) \approx 1$ which gives the fixed-point method as a special case of the Newton-Raphson scheme.

***Iteration 3:***

New value $x_3$ is again evaluated in the same way using Equation (15):

$$x_3 = x_2 - \frac{F(x_2)}{F'(x_2)} \tag{15}$$

In Equation (15), $F'(x_2)$ can be calculated or $F'(x_1)$ or $F'(x_0)$ can be reused. Also, $F(x_2)$ can be calculated using $z_{02} = \frac{y_0}{y_2}$, where $y_2 = \frac{2.51 \cdot x_2}{Re} + \frac{\varepsilon^*}{3.71}$. Input parameter for Padé polynomial $z_{02}$ here refers to $y_0$ from the first iteration; Equation (16). It can be evaluated also using $z_{12} = \frac{y_1}{y_2}$, always with the reference to the preceding iteration (here to the second iteration); Equation (17).

$$F(x_2) = x_2 + 2 \cdot log_{10}(y_0) - \frac{(z_{02} - 1) \cdot (11 \cdot z_{02}^2 + 38 \cdot z_{02} + 11)}{2.302585093 \cdot (3 \cdot z_{02}^3 + 9 \cdot z_{02}^2 + 9 \cdot z_{02} + 1)} \tag{16}$$

In Equation (16), $log_{10}(y_0) - \frac{(z_{02}-1)\cdot(11\cdot z_{02}^2+38\cdot z_{02}+11)}{2.302585093\cdot(3\cdot z_{02}^3+9\cdot z_{02}^2+9\cdot z_{02}+1)} \equiv log_{10}(y_2)$.

The Padé polynomial is a very accurate approximation of logarithmic function, so knowing that $y_0$ is evaluated directly through the logarithmic function, while $y_1$, $y_2$, $y_3$, etc. is based on its Padé polynomial equivalent, it is obvious that the sequence $z_{01} = \frac{y_0}{y_1}$, $z_{02} = \frac{y_0}{y_2}$, $z_{03} = \frac{y_0}{y_3}$, etc. is slightly more accurate compared with the sequence $z_{01} = \frac{y_0}{y_1}$, $z_{12} = \frac{y_1}{y_2}$, $z_{23} = \frac{y_2}{y_3}$, etc. which accumulates error introduced with Padé approximations. On the other hand, the error is minimized when the argument $z$ is closer to 1 which is case for the second sequence $z_{01} = \frac{y_0}{y_1}$, $z_{12} = \frac{y_1}{y_2}$, $z_{23} = \frac{y_2}{y_3}$, etc. In both cases, the introduced error can be neglected.

$$F(x_2) = x_2 + 2\cdot log_{10}(y_1) - \frac{(z_{12}-1)\cdot(11\cdot z_{12}^2+38\cdot z_{12}+11)}{2.302585093\cdot(3\cdot z_{12}^3+9\cdot z_{12}^2+9\cdot z_{12}+1)} =$$
$$x_2 + 2\cdot log_{10}(y_0) - \frac{(z_{01}-1)\cdot(11\cdot z_{01}^2+38\cdot z_{01}+11)}{2.302585093\cdot(3\cdot z_{01}^3+9\cdot z_{01}^2+9\cdot z_{01}+1)} - \frac{(z_{12}-1)\cdot(11\cdot z_{12}^2+38\cdot z_{12}+11)}{2.302585093\cdot(3\cdot z_{12}^3+9\cdot z_{12}^2+9\cdot z_{12}+1)}$$
(17)

In Equation (17), $log_{10}(y_1) - \frac{(z_{12}-1)\cdot(11\cdot z_{12}^2+38\cdot z_{12}+11)}{2.302585093\cdot(3\cdot z_{12}^3+9\cdot z_{12}^2+9\cdot z_{12}+1)} \equiv log_{10}(y_2)$ and,

$log_{10}(y_0) - \frac{(z_{01}-1)\cdot(11\cdot z_{01}^2+38\cdot z_{01}+11)}{2.302585093\cdot(3\cdot z_{01}^3+9\cdot z_{01}^2+9\cdot z_{01}+1)} \equiv log_{10}(y_1)$

***Iteration $i$:***

All indexes $i$ in respect the third iteration should be updated as $i = i + 1$ with exemption of index 0 in Equation (16). The calculation is finished when $x_{i+1} \approx x_i$.

The algorithm for the proposed improved procedure is given in Figure 2.

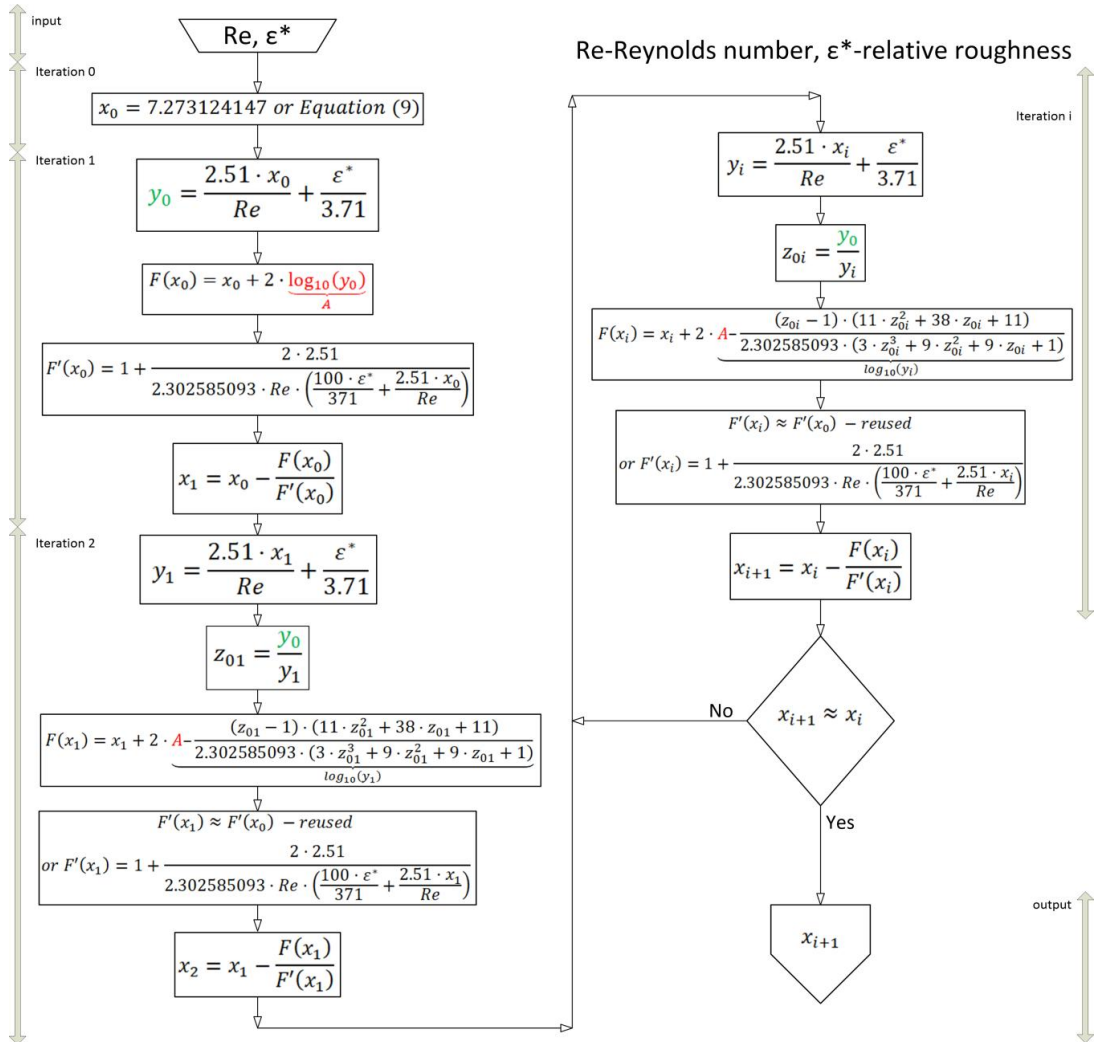

Figure 2. Algorithm for the proposed one *log*-call improved procedure.

Only a one-off evaluation of the logarithmic function is needed in the proposed algorithm from Figure 2, which is clearly marked in red; $A = \log_{10}(y_0)$. On the other hand, $y_0$ calculated in iteration 1 is reused in all next steps and it is marked in green in Figure 2.

The proposed procedure can be simplified assuming that $F'(x_i) = 1$ which gives the simple fixed-point procedure (Brkić 2017b) instead of the Newton-Raphson.

## 5. Numerical examples

Here are shown two numerical examples:

Example 1:                                              Example 2:
$Re = 8.31 \cdot 10^3, \ \varepsilon^* = 0.024$          $Re = 2.5 \cdot 10^6, \ \varepsilon^* = 4 \cdot 10^{-4}$

Iteration 0
$x_0 = 6.279860788$                                      $x_0 = 7.401979091$                          (9)

Iteration 1
$y_0 = 0.008365808$                                      $y_0 = 0.000115248$
$\log_{10}(y_0) = -2.077492116$                          $\log_{10}(y_0) = -3.938365477$
$F(x_0) = 2.124876556$                                   $F(x_0) = -0.474751864$                      (11)
$F'(x_0) = 1.001337518$                                  $F'(x_0) = 1.001024781$                      (12)
$x_1 = 4.157822498$                                      $x_1 = 7.876244936$                          (10)

Iteration 2
$y_1 = 0.007724855$                                      $y_1 = 0.000115724$
$z_{01} = \frac{y_0}{y_1} = 1.082972765$                 $z_{01} = \frac{y_0}{y_1} = 0.995885374$
0.034617535                                              -0.001790646        Padé approximant (6)
$F(x_1) = -0.066396805$                                  $F(x_1) = 0.003095273$                      (14)
$F'(x_1) = 1.001986711$                                  $F'(x_1) = 1.000970478$
$x_2 = 4.224087653$                                      $x_2 = 7.873152664$                          (13)

Iteration 3
$y_2 = 0.00774487$                                       $y_2 = 0.000115721$
$z_{02} = \frac{y_0}{y_2} = 1.080174034$                 $z_{02} = \frac{y_0}{y_2} = 0.995912092$
0.033493733                                              -0.001778995        Padé approximant (6)
$F(x_2) = 0.002115955$                                   $F(x_2) = -2.03017 \cdot 10^{-5}$           (16)
$F'(x_2) = 1.001957048$                                  $F'(x_2) = 1.000970813$
$x_3 = 4.221975832$                                      $x_3 = 7.873172946$                          (15)

Final value:
$x = 4.22204103$                                         $x = 7.873172814$

## 6. Conclusions

An efficient algorithm for the iterative calculation of the Colebrook equation by both an accurate and computationally efficient Padé approximation is presented in this paper. It requires only one evaluation of the logarithmic function in respect to the whole iterative procedure and more specifically only in the first iteration, while the common procedures from current engineering practice require at least one evaluation of logarithmic function for every single iteration. The logarithmic function in the proposed procedure is replaced in all iterations (except the first), with simple, accurate and efficient Padé polynomials (Baker and Graves-Morris 1996). In this way the same accuracy is reached through the proposed less demanding procedure, after the same number of iterations as in the standard algorithm which uses *log*-call in each iterative step. This is a good achievement knowing that the nature of the Colebrook equation is logarithmic. For their evaluation in the Central Processor Unit (CPU) of computers, Padé polynomials require a lower number of floating-point operations to be executed compared with the logarithmic function (Clamond 2009, Giustolisi et al. 2011, Danish et al. 2011, Winning and Coole 2013, Vatankhah 2018, Sonnad and Goudar 2004, Brkić 2012a, Winning and Coole 2015).

The here presented iterative approach only introduces a computationally cheaper alternative to the standard iterative procedure. It does not reduce the number of required iterations to reach the final desired accuracy nor provide more accurate results. The proposed method reduces the burden for the Central Processing Unit (CPU) as less floating-point operations need to be executed. In that way, the presented approach also increases speed of computation. On the other hand, many explicit non-iterative approximations to the Colebrook equation are available in literature (Gregory and Fogarasi 1985, Zigrang and Sylvester 1985, Brkić 2011e, Pimenta et al. 2018) which initially appear simple for computation, but are not. They are widely used, but although some of them are very accurate, they contain relatively complex internal iterative steps and also a number of computationally demanding functions. For example, the widely used Haaland approximation introduces relative error up to 1.5% (Haaland 1983, Wood and Haaland 1983), but with the cost of evaluation of one logarithmic expression and one non-integer power. Also, the approximation by Romeo et al. (2002) reaches extremely high accuracy with the relative error of up to 0.14%, but with a cost of evaluation of even three logarithmic expressions and two non-integer powers. Regarding alternative iterative procedures, Clamond (2009) provides an accurate iterative approach using $\Omega$ function, but this algorithm requires at least two *log*-calls; one for initialization and one in the first iteration, which is more expensive compared with the here presented approach.

The procedure proposed in this paper can significantly reduce the computational burden for evaluating complex distribution networks with various applications (water, gas) (Brkić 2009, Brkić 2011a, Praks et al. 2015, Praks et al. 2017, Brkić 2016, Brkić 2018). For example, a probabilistic approach using time dependent modeling of distribution or transmission networks requires many millions of evaluations of Colebrook's equation, which means that it is not a computationally cheap task at all. For such kinds of computations is always good to have a cheaper but still accurate approach in order to speed up the process.

**Acknowledgments:** We thank Adrian O'Connell who as native speaker kindly checked correctness of English expression through the paper.

**Conflicts of Interest:** The authors declare no conflict of interest. The views expressed in this paper are those of the authors.

**Further reading**